\documentclass[aps,twocolumn,amsmath,amssymb,prl]{revtex4-2}

\usepackage{setspace}
\usepackage{wasysym}
\usepackage{epsfig}
\usepackage{epstopdf} 
\usepackage{pstricks}
\usepackage{multirow}
\usepackage{booktabs}
\usepackage{array}
\usepackage{mwe}
\usepackage{soul}

\def\be{\beta}
\def\ga{\gamma}

\def\ve{\varepsilon}

\def\et{\eta}

\def\la{\lambda}

\def\si{\sigma}

\def\ch{\chi}
\def\ps{\psi}
\def\om{\omega}

\def\La{\Lambda}
\def\Si{\Sigma}

\def\Om{\Omega}
\def\mn{{\mu\nu}}

\def\cL{{\cal L}}
\def\lrvec#1{ \stackrel{\leftrightarrow}{#1} }

\def\half{{\textstyle{1\over 2}}}
\def\quar{{\textstyle{1\over 4}}}

\def\frac#1#2{{\textstyle{{#1}\over {#2}}}}

\def\prt{\partial}

\def\ol#1{\overline{#1}}

\newcommand{\beq}{\begin{equation}}
\newcommand{\eeq}{\end{equation}}
\newcommand{\bea}{\begin{eqnarray}}
\newcommand{\eea}{\end{eqnarray}}
\newcommand{\bit}{\begin{itemize}}
\newcommand{\eit}{\end{itemize}}
\newcommand{\rf}[1]{(\ref{#1})}

\def\bt{{\tilde b}}

\def\dt{{\tilde d}}
\def\gt{{\tilde g}}
\def\Ht{{\tilde H}}

\begin{document}
\raggedbottom

\title{Complete-Coverage Searches for Lorentz Violation
in the Minimal Matter Sector}

\author{Marshall J. Basson}
\altaffiliation[Present address: ]{Department of Physics and Astronomy, Michigan State University, East Lansing,
Michigan 48824, USA}
\author{Eric Biddulph-West}
\author{Caitlyn Holl}
\author{Will Lankenau}
\author{Facundo Martin Lopez}
\author{Bianca Rose Lott}
\author{Chihui Shao}
\author{Danny P. Shope}
\author{Jay D. Tasson}
\email[Corresponding author: ]{jtasson@carleton.edu}
\author{Zhiyu Zhang}
\affiliation{Physics and Astronomy Department, Carleton College, Northfield, MN 55057, USA}

\date{August 2026}

\begin{abstract}
Over the past several decades, dozens of tests have sought 
the 132 Lorentz-violating degrees of freedom
in the nonrelativistic limit of the minimal matter sector of the Standard-Model Extension,
yet 43 remained unconstrained.
In this Letter, 
we limit all previously unconstrained degrees of freedom and make improvements on 13 prior limits. 
The approach introduced here offers the potential of further improvements for 49 degrees of freedom in suitable future experiments,
along with additional discovery potential offered by combining data from experiments performed in different locations.
\end{abstract}

\maketitle

{\it Introduction---}
Several decades ago,
the identification of Lorentz violation
as a possible signal of new physics at the Planck scale
\cite{ksstring89,*kpstrings91}
initiated a renewed interest in tests of Lorentz symmetry.
A framework for organizing a systematic search
for Lorentz violation across all of physics
has been developed
and is known as the Standard-Model Extension (SME) \cite{ck97,*ck98,*k04}.
Lorentz violation can be triggered by various mechanisms \cite{bluhm19,*bluhm05},
but generally can be modeled via a condensate of background vectors and tensors
called coefficients for Lorentz violation.
The coefficients provide preferred directions in empty spacetime
and hence generate violations of rotation and boost invariance
for systems involving particles that couple to the backgrounds, resulting in broken global Lorentz symmetry \cite{akzl}.

The SME is constructed as an effective field theory
that provides an expansion about known physics
in operators of increasing mass dimension.
The additional terms are constructed from the known fields
in the Standard Model and General Relativity
along with the coefficients for Lorentz violation.
The terms in this expansion with the same mass dimension
as the operators in known physics
are known as the minimal SME \cite{[{For a review addressing these points, see, }]tasson14}.

Because of the relative simplicity of the terms involved
in the minimal SME,
and the fact that precision experiments involving ordinary matter
are relatively common,
dozens of searches for coefficients for Lorentz violation 
in the minimal SME associated
with ordinary matter (protons, neutrons, and electrons) 
in flat spacetime have been performed.
These tests have now reached truly impressive levels of sensitivity.
For example,
some coefficients with units of mass
are limited to less than $10^{-29}$ times the mass of the electron,
which by conventional thinking is
well beyond one Planck suppression.
Despite this effort,
a significant number of coefficients in this sector remain unconstrained \cite{datatable}.

In nonrelativistic studies involving protons, neutrons, electrons, and their antiparticles, there are 132 independently observable coefficients for Lorentz violation, among which 43 remain unconstrained \cite{datatable}.
In this Letter,
we provide a complete approach to addressing this deficiency,
which utilizes existing high-precision experiments
analyzed to higher order in their boost velocities.
In addition to providing 43 new constraints,
we improve existing limits on 13 coefficients.
Setting these limits is the first of three key results presented in this Letter.
Our methods also form a template for improving sensitivities to 49 coefficients in future experiments,
and we provide estimates for the level of sensitivity that might be obtained in such efforts.
Finally, this Letter shows that independent access to the full set of minimal matter-sector coefficients considered together in a coefficient separation approach can be obtained.  These latter results offer the opportunity for significantly expanded discovery potential.

{\it Background---}
The Lagrange density for the minimal matter sector of the SME
takes the form \cite{ck98}
\bea
\nonumber
\cL &=& \frac{1}{2} i \overline{\ps} 
(\ga_\nu + c_\mn \ga^\mu + d_\mn \ga_5 \ga^\mu 
   + \half g_{\la \mu \nu} \si^{\la \mu})
\lrvec{\prt^\nu} \ps \\
& &
- \ol{\ps} 
(m + b_\mu \ga_5 \ga^\mu 
   + \half H_\mn \si^\mn)
   \ps.
\label{lagr}
\eea
Here $\ps$ is the fermion field,
$\ga^\mu$ denotes the standard Dirac matrices,
$m$ is the conventional Lorentz-invariant mass,
and the objects 
$b_\mu$, $c_\mn$, $d_\mn$, $g_{\la\mn}$, and $H_\mn$ 
are the coefficients for Lorentz violation.
For brevity,
we omit coefficients $a_\mu$, $e_\mu$, and $f_\mu$,
which can be removed from the theory in the limit considered here \cite{ck98, Altschulf}.
We avoid further discussion of $c_\mn$
because all of its components have been probed already
by existing analysis \cite{datatable}.
We also omit the trace of $d_\mn$ from consideration
as it is not Lorentz violating.
The coefficients have no special symmetries
beyond the defining antisymmetry of 
$g_{\la\mn}$ and $H_\mn$ on the first two indices.
The 44 observable degrees of freedom per fermion
are chosen as combinations of the coefficients above
denoted with a related letter and an overtilde,
the so-called ``tilde coefficients'' \cite{datatable}.
For example,
the combination $\bt_J$,
which will play a key role in this work,
takes the form
\beq
\bt_J = b_J -\half\ve_{JKL}H_{KL}-m(d_{JT}-\half\ve_{JKL}g_{KLT}).
\label{sunbt}
\eeq
The complete set of tilde-coefficient definitions,
as well as the inverse transformations
can be found in Tables P56 and P57 of Ref.~\cite{datatable}.

By convention,
the coefficients for Lorentz violation
are taken
as spacetime constants
in Cartesian coordinates in a standard Sun-centered reference frame,
which is inertial to a good approximation
over the timescale of the relevant experiments \cite{datatable}.
The coefficients are then time dependent
in the frame of Earth-based laboratories
due to various motions that include
the sidereal rotation of the laboratory,
the time-dependent boost of Earth as it revolves around the Sun ($\be_\oplus \sim 10^{-4}$) at the annual frequency $\Om$, and
the time-dependent boost of the laboratory as it revolves around Earth's rotation axis ($\be_L\sim10^{-6}$) at the sidereal frequency $\om_L$.
In the presence of Lorentz violation,
these motions typically generate periodic signals
in laboratory systems at the associated annual and sidereal frequency and harmonics and combinations of them, which form the potential experimental signatures of Lorentz violation that we exploit.
In this Letter,
we denote components of vectors and tensors in the Sun-centered frame
with capital indices and components in the laboratory frame with lower-case indices. 
Nonzero values for the coefficients for Lorentz violation
generally result in violations of both boost and rotation invariance.
It is only in special limits of the SME where
special frames can be found in which rotation invariance is preserved.

{\it Extended frame transformations---} 
When signals are sought via boost effects,
they are suppressed compared to rotation-invariance tests by multiples of the boost velocities involved.
For this reason,
along with the complicated nature of the frame transformations,
most tests
have focused on signals associated with Earth's rotation
or one power of $\be_\oplus$ (see, e.g., Refs.~\cite{lopez23, Sanner2019, fgt17, expdata:electron-washington, Matveev2013, *CBP2004, *Altschul2010, *KosVar2015, *KosVar2018}). 
To date only two analyses have considered effects at second order in the boost velocities: one in the context of the $c_\mn$ coefficients in the matter sector \cite{cbsquare}
and the other in the photon sector \cite{kmm16}.
In this Letter, we exploit effects 
that arise with suppression factors of up to 20 orders of magnitude coming from one boost suppression by $\be_\oplus$ or $\be_L$ as well as by combinations arising at 
second 
\{$\be_\oplus^2, \be_L \be_\oplus, \be_L^2$\}, 
third \{$\be_\oplus^3, \be_L \be_\oplus^2,\be_L^2 \be_\oplus, \be_L^3$\}, fourth \{$\be_\oplus^4,\be_L \be_\oplus^3, \be_L^2 \be_\oplus^2$\}, and fifth \{$\be_\oplus^5$\} order in the boost factors.
Note that combinations at fourth and fifth order that would generate a suppression by more than 20 orders of magnitude are not considered here.
Though a circular model for Earth's orbit is sufficient for most purposes, we use an eccentric model of a Kepler orbit here taking into consideration the time-dependent variations of $\beta_\oplus$.
The eccentricity of Earth's orbit $\varepsilon \sim 10^{-2}$ can be treated as an additional small quantity in our perturbative analysis
generating an additional set of suppression factors that can be obtained from the above lists via the replacement $\be_\oplus^p \rightarrow \varepsilon^{2q} \be_\oplus^{p-q}$ for integer powers $p>1$ and integers $1\leq q<p$. Our work takes into account the precession of the Earth's spin axis since the year 2000, the time origin of the standard coordinate system for SME studies.  This is relevant because experiments seeking sidereal effects performed at different times are sensitive to different linear combinations of coefficients.  We also consider the impact of Earth's equatorial bulge on $\be_L$, but find no significant effects.

We choose to end our perturbation series at 20 orders of suppression as it allows the inclusion of terms that are necessary and sufficient to provide complete coverage in the matter sector. 
Further,
when applied to the most sensitive experiments,
limits emerge
that are competitive with other existing limits on Lorentz violation.

At the level of the nonrelativistic Hamiltonian
relevant for laboratory experiments that results from
Eq.~\rf{lagr},
a number of terms appear that can be useful in probing Lorentz violation \cite{kl99}.
Here,
we focus on the term 
\beq
H \supset H_b = \bt_j(t) \si^j,
\eeq
where $\bt_j(t)$ is the laboratory-frame version of $\bt_J$
and $\si^j$ are the Pauli matrices.
While consideration of additional terms in $H$
is of definite interest,
we focus on $H_b$ due to the high sensitivities to this term that have been achieved \cite{expdata:electron-washington,expdata:neutron-princeton,expdata:neutron-berlin},
the large number of relevant experiments that have been done \cite{datatable},
and the breadth and depth of sensitivities that result from the current study
when this term is considered.

Since $\bt_J$ is a vector under observer rotations,
$\bt_j(t)$ can be achieved from $\bt_J$ via the application
of a simple rotation matrix
when boosts are ignored.
To achieve the form of $\bt_j(t)$ to arbitrary order in the boost parameters,
we proceed by applying the full transformation from the Sun-centered frame
to the laboratory frame
to each of the coefficients \{$b_J,d_{TJ},H_{KL},g_{KLT}$\}.
For a coefficient with a generic number of indices $t_{\Si \Xi \ldots}$
this transformation is
\beq
t_{\si \xi \ldots} = \La^\Si_{\phantom{\Si} \si} \La^\Xi_{\phantom{\Xi} \xi} t_{\Si \Xi \ldots},
\eeq
where $\La^\Si_{\phantom{\Si} \si}$ is the full transformation matrix
including a boost from the Sun-centered frame
to the Earth-centered frame, a boost from the Earth-centered frame
to a frame traveling with the laboratory on the surface of Earth,
and finally a rotation that aligns the coordinates with the standard fixed choice in the laboratory \cite{km02}.
The results of this transformation applied to each coefficient are then inserted into the laboratory-frame definition of $\bt_j(t)$
to achieve its form in terms of the constant Sun-centered frame coefficients and its explicit time dependence.
The transformation to tilde coefficients can then be used
to write the Lorentz-violating degrees of freedom
in the standard form used in the literature. 
This approach allows us to
extend the existing work, which treats experiments as occurring in instantaneous Lorentz frames within special relativity (see, e.g., Ref.~\cite{Bluhm_2003}), to arbitrary powers in a series expansion in the relevant boost parameters. 
The consideration of noninertial frame effects as well as the effects of gravity may also be of interest, but are beyond our present scope \cite{[{For related efforts in gravity, see, }] kt09,*kt11,*bonder,*kl21,*colladay19}.

{\it Constraints and suggested searches---} 
A given experiment can potentially make a large number of independent measurements
by seeking each of the relevant Fourier components of $\bt_j(t)$
for each of the three Cartesian components in the laboratory.
At the order to which we work, signals can be found at frequencies 
\beq
\om = n_1 \om_L \pm n_2 \Om
\label{freq}
\eeq
for $n_1 \in \{0,1\}$
and $n_2 \in \{0,1,2,3,4,5,6,7,8,9\}$, or $n_1 \in \{2\}$
and $n_2 \in \{0,1,2,3,4,5,6\}$.
Moreover,
the amplitude of a signal
typically depends on the colatitude at which the experiment is performed.
Hence,
experiments at different colatitudes probe different linear combinations of coefficients. As one example of the structures that arise, which are, in general, quite lengthy,
we present explicitly the leading contributions to $\bt_x$ that vary as $\sin{(2\om_L t+\Om T)}$ in a simpler circular orbit model that is independent of the precession of Earth's spin axis,
\bea
\nonumber
\bt_x &\supset& \quar \be_L \be_\oplus \cos(\chi)[\sin^2({\textstyle{\eta\over{2}}})(\bt_Y - 2 \dt_Y + \gt_{DY} + \gt_{TY})\\
& &
    +2 \sin({\textstyle{\eta\over{2}}})\cos({\textstyle{\eta\over{2}}}) \gt_{TZ} 
    ]\sin(2\om_L t + \Om T).
\eea
Here $T$ is the time coordinate of the Sun-centered frame, $t$ is measured from one of the times at which the local coordinates align with the Sun-centered frame coordinates,
$\et$ is the inclination of Earth's orbit,
$\ch$ is the colatitude of the laboratory,
and $\dt_Y$, $\gt_{DY}$, $\gt_{TY}$, and $\gt_{TZ}$
are additional tilde coefficients that emerge as observables in this double boost-suppressed contribution.

In the context of a test framework such as the SME
that is broad and general so as to capture the physical effects
of many possible models,
one must decide how many degrees of freedom to fit at one time.
One popular approach is to consider the coefficients one
Lorentz-violating degree of freedom at a time
in a so-called maximum-reach analysis \cite{fgt17}.
The other approach, known as a coefficient separation approach, is to consider some larger subset,
such as all of the components of a single coefficient,
or all of the minimal coefficients associated with a particular particle,
etc.\ \cite{fgt17}.
In this Letter,
we present results in the context of a maximum-reach analysis.
However, our methods enable exciting new possibilities in the context of a coefficient separation approach as well.

We perform a maximum-reach analysis for all three species of ordinary matter using experimental data from
Ref.~\cite{expdata:neutron-berlin} for the neutron; Ref.~\cite{expdata:electron-washington} for the electron; and Ref.~\cite{expdata:neutron-berlin}, as adapted in Ref.~\cite{expdata:proton-conversion}, for the proton. Each of these experiments used $H_b$ to measure some combination of $\bt_{x,y}(t)$ at the sidereal frequency to extract limits on $\bt_J$.
In this Letter,
we extract real
maximum-reach bounds appearing as bold numbers in the ``Current" columns of Table~\ref{datatables}. 
These amount to 43 new first-ever maximum-reach limits on coefficients (boxed), as well as 13 improvements on previous bounds. 

For coefficients that appear at frequencies other than the sidereal frequency
that have not been studied experimentally,
we provide estimated sensitivities based on the assumption that the appropriate amplitudes
could be extracted with an intrinsic experimental sensitivity similar to existing results at the sidereal frequency. Order-of-magnitude estimates for the best sensitivities that can be achieved for the relevant coefficients using the methods presented here are reported in the ``Feasible" columns of Table~\ref{datatables} along with a frequency at which they can be found. In total, this results in 67 coefficients that we either constrain for the first time, provide an improved bound, or discover the potential for improvement via an analysis at a suitable additional frequency.

Note that when near-sidereal frequencies are used,
care must be taken to decide if these signals are sufficiently close to the sidereal frequency
such that they would be detected in a sidereal search,
or if they are separable and require specific consideration of these Fourier components.
The former perhaps offers some simplicity in extracting maximum-reach bounds
while the latter provides more independent measurements useful
in simultaneously fitting multiple coefficients. 
These ideas have already been exploited for one suppression by $\be_\oplus$
using the frequencies $\om_L \pm 2 \Om$ in electron experiments \cite{Sanner2019}.
It is also possible to deal with partially overlapping Fourier components 
\cite{Chung2009}.
When similar sensitivities to a given coefficient could be achieved at multiple different frequencies, Table~\ref{datatables} presents the frequency that is likely to be the easiest to separately observe.

Perhaps even more interesting than extracting the new limits
is the additional discovery potential provided by studies at higher powers of the boost. This potential arises from two basic sources:
seeking time dependence at additional frequencies
and performing experiments at different latitudes.
Consider first the additional frequencies in Eq.~\rf{freq}.
The estimated sensitivities in Table~\ref{datatables} are arising
because additional coefficients are appearing at these additional frequencies
that have not yet been explored in the data.
Hence exploring these frequencies offers discovery potential.
Similarly, $\be_L$
is a function of latitude---experiments closer to the equator
experience a higher boost than those near the poles. In some cases additional latitude dependence arises due to experimental considerations associated with the alignment of the experiment with the local vertical direction.
Hence, otherwise identical experiments at different latitudes
can measure linearly independent combinations of coefficients. 
Fundamentally, this implies that Lorentz violation can evade detection
in an experiment at one latitude while being detected in an experiment at another latitude.

\newcolumntype{M}[1]{>{\centering\arraybackslash}m{#1}}
\begin{table*}[tbph]
\begin{tabular}{ M{0.8cm} ||M{1.4cm} |M{1.4cm} M{1.6cm} || M{1.4cm} |M{1.4cm} M{1.6cm} || M{1.4cm} |M{1.4cm} M{1.6cm} }
\hline
\multirow{2}{*}{Coeff} &
  \multicolumn{3}{c|}{Electron} &
  \multicolumn{3}{c|}{Proton} &
  \multicolumn{3}{c}{Neutron} \\ \cline{2-10} 
 &
  \multicolumn{1}{c|}{Current} &
  \multicolumn{1}{c|}{Feasible} &
  Frequency &
  \multicolumn{1}{c|}{Current} &
  \multicolumn{1}{c|}{Feasible} &
  Frequency &
  \multicolumn{1}{c|}{Current} &
  \multicolumn{1}{c|}{Feasible} &
  Frequency \\ \hline\hline
$\bt_{X} $& {$10^{-31}$} & - & - &      {$10^{-32}$} & - & - &       {$10^{-33}$} & - & - \\
$\bt_{Y} $& {$10^{-31}$} & - & -&        {$10^{-32}$} & - & - &       {$10^{-33}$} & - & - \\
$\bt_{Z} $&
  $10^{-29}$ &
  - &
  -&
  ${\bf 10^{-29}}$ &
   - &
  -
   &
  {$\bf10^{-30}$} &
  - &
  -
   \\
$\bt_{T} $&
  {$10^{-26}$} &
   - &
  - &
  $\bf10^{-26}$ &
  $\bf10^{-28}$ &
  $\om_L+\Om$
   &
  {$\bf10^{-27}$} &
  $\bf10^{-29}$ &
  $\om_L+\Om$ 
  \\
  $\bt^*_{X} $&
  {$10^{-22}$} &
   - &
  -&
  $10^{-24}$ &
  - &
  -
   &
  {$\boxed{\bf 10^{-12}}$} &
  {$\boxed{\bf10^{-16}}$} &
  $\om_L-2\Om$
  \\
  $\bt^*_{Y} $&
  {$10^{-22}$} &
   - &
  -&
  $10^{-24}$ &
  - &
  -
   &
  {$\boxed{\bf10^{-15}}$} &
  {$\boxed{\bf10^{-16}}$} &
  $\om_L-2\Om$ 
  \\
  $\bt^*_{Z} $&
  {$10^{-23}$} &
   - &
  -&
  $10^{-24}$ &
  - &
  -
   &
  {$\boxed{\bf10^{-15}}$} &
  - &
  - 
   \\[.5cm]

$\dt_{+} $&
  {$10^{-27}$} &
  - &
  - &
  ${\bf 10^{-27}}$ &
   {${\bf 10^{-28}}$} &
  {$\Om$}
   &
  {$10^{-27}$} &
   {${\bf 10^{-29}}$} &
  $\Om$
   \\
$\dt_{-} $&
  {$10^{-26}$} &
  {-} &
  - &
 {$\boxed{\bf{10^{-26}}}$} &
 $\boxed{\bf10^{-28}}$ &
  $\om_L+\Om$
   &
  {$\bf10^{-27}$} &
   {$\bf10^{-29}$} &
  {$\om_L+\Om$}
   \\
$\dt_{Q} $&
  {$10^{-26}$} &
  {-} &
  - &
  {$\bf 10^{-26}$} &
   {$\bf10^{-28}$} &
  {$\Om$}
   &
  {$\bf10^{-27}$} &
  {$\bf10^{-29}$} &
  $\om_L-\Om$
   \\
$\dt_{XY} $&
  {$10^{-26}$} &
    {-} &
  - &
  {$\boxed{\bf 10^{-26}}$} &
   {$\boxed{\bf10^{-28}}$} &
  {$\om_L+\Om$}
   &
  {$10^{-27}$} &
  {$\bf10^{-29}$} &
  $\om_L+\Om$
   \\
$\dt_{YZ} $&
  {$10^{-26}$} &
  - &
  - &
  $\boxed{\bf 10^{-26}}$ &
   {$\boxed{\bf10^{-27}}$} &
  {$\om_L+\Om$}
   &
  {$\bf10^{-27}$} &
  {$\bf10^{-28}$} &
  $\om_L+\Om$
   \\
$\dt_{ZX} $&
  {$10^{-26}$} &
   {-} &
   - &
  {$\boxed{\bf{10^{-26}}}$} &
  {$\boxed{\bf10^{-28}}$} &
  $\Om$
   &
  {$\boxed{\bf{10^{-27}}}$} &
  {$\boxed{\bf10^{-29}}$} &
  $\Om$
   \\
$\dt_{X} $&
  {$10^{-22}$} &
  - &
  -&
  {$10^{-27}$} &
  - &
  -
   &
  {$10^{-28}$} &
  - &
  -
   \\
$\dt_{Y} $&
  {$10^{-22}$} &
  - &
  - &
  {$10^{-27}$} &
   - &
   -
   &
  {$10^{-28}$} &
  - &
  -
   \\
$\dt_{Z} $&
  {$\bf 10^{-22}$} &
  {-} &
  - &
  {$\boxed{\bf{10^{-23}}}$} &
  - &
  -
   &
  {$\boxed{\bf{10^{-24}}}$} &
  - &
  -
   \\ [.5cm]
$\Ht_{XT} $&
  {$10^{-26}$} &
  {-} &
  - &
  $\boxed{\bf{10^{-26}}}$ &
  {$\boxed{\bf10^{-28}}$} &
  $\Om$
   &
  {$\bf10^{-27}$} &
  {$\bf10^{-29}$} &
  $\Om$
   \\ 
$\Ht_{YT} $&
  {$10^{-26}$} &
  {-} &
  -&
  {$\boxed{\bf{10^{-26}}}$} &
  {$\boxed{\bf10^{-28}}$} &
  $\Om$
   &
  {$\bf{10^{-27}}$} &
  {$\bf 10^{-29}$} &
  $\Om$
   \\ 
$\Ht_{ZT} $&
  {$10^{-26}$} &
  {$\bf10^{-27}$} &
  $\om_L-\Om$&
  {$\boxed{\bf{10^{-26}}}$} &
  {$\boxed{\bf10^{-28}}$} &
  $\om_L-\Om$
   &
  {$10^{-27}$} &
  {$\bf10^{-29}$} &
  $\om_L-\Om$
   \\ [.5cm]
$\gt_{T} $&
  $10^{-27}$ &
  - &
  - &
  $\bf10^{-26}$ &
  {$\bf10^{-28}$} &
  $\Om$ &
  $10^{-27}$ &
  $\bf10^{-29}$ &
  $\om_L+\Om$ 
   \\ 
$\gt_{c} $&
  $10^{-26}$ &
  $\bf10^{-27}$ &
  $\om_L+\Om$ &
  {$\boxed{\bf10^{-26}}$} &
  {$\boxed{\bf10^{-28}}$} &
  $\om_L+\Om$ &
  $10^{-27}$ &
  $\bf10^{-29}$ &
  $\om_L+\Om$
   \\ 
$\gt_{Q} $&
  {$\boxed{\bf10^{-21}}$}  &
  {-} &
  -&
  {$\boxed{\bf10^{-23}}$}  &
  - &
  - &
  {$\boxed{\bf10^{-24}}$}  &
  - &
  -
   \\ 
$\gt_- $&
  {$\boxed{\bf10^{-21}}$}  &{$\boxed{\bf10^{-22}}$} &
  $2\Om$ &
  {$\boxed{\bf10^{-23}}$}  &
  {$\boxed{\bf10^{-24}}$} &
  $2\Om$ &
  {$\boxed{\bf10^{-24}}$} &
  - &
  -
   \\ 
$\gt_{TX} $&
  {$\boxed{\bf10^{-22}}$}  &
  - &
  - &
  {$\boxed{\bf10^{-23}}$}  &
  - &
  - &
 {$\boxed{\bf10^{-24}}$}  &
  - &
  -
   \\ 
$\gt_{TY} $&
  {$\boxed{\bf10^{-22}}$}  &
  - &
  - &
  {$\boxed{\bf10^{-23}}$}  &
  {$\boxed{\bf10^{-24}}$} &
  $\om_L-2\Om$ &
  {$\boxed{\bf10^{-24}}$}  &
  - &
    -   \\ 
$\gt_{TZ} $&
  {$\boxed{\bf10^{-21}}$}  &
{$\boxed{\bf10^{-22}}$} &
  $2\Om$&
  {$\boxed{\bf10^{-23}}$}  &
  {$\boxed{\bf10^{-24}}$} &
  $2\Om$ &
  {$\boxed{\bf10^{-24}}$}  &
  - &
  -
   \\ 
$\gt_{XY} $&
  $10^{-17}$ &
  {$\bf10^{-18}$} &
  $\Om$ &
  {$\boxed{\bf10^{-18}}$} &
  {$\boxed{\bf10^{-20}}$} &
  $\Om$ &
  {$\boxed{\bf10^{-18}}$} &
  {$\boxed{\bf10^{-20}}$} &
  $\Om$
   \\ 
$\gt_{YX} $&
  $10^{-17}$ &
  {$\bf10^{-18}$} &
  $\Om$ &
  {$\boxed{\bf10^{-18}}$} &
  {$\boxed{\bf10^{-20}}$} &
  $\Om$ &
  {$\boxed{\bf10^{-19}}$} &
  {$\boxed{\bf10^{-20}}$} &
  $\Om$
   \\ 
$\gt_{ZX} $&
  $10^{-18}$ &
  {-} &
  -&
  {$\boxed{\bf10^{-18}}$} &
  {$\boxed{\bf10^{-20}}$} &
  $\om_L-\Om$ &
  {$\boxed{\bf10^{-19}}$} &
  {$\boxed{\bf10^{-20}}$} &
  {$3\Om$}
   \\ 
$\gt_{XZ} $&
  $10^{-17}$ &
  {$\bf10^{-18}$} &
  $\om_L-3\Om$ &
  {$\boxed{\bf10^{-18}}$} &
  {$\boxed{\bf10^{-19}}$} &
  {$3\Om$} &
  {$\boxed{\bf10^{-18}}$} &
  {$\boxed{\bf10^{-20}}$} &
  $\Om$
   \\ 
$\gt_{YZ} $&
  $10^{-17}$ &
  {$\bf10^{-18}$} &
  $\om_L-3\Om$ &
  {$\boxed{\bf10^{-17}}$} &
  {$\boxed{\bf10^{-19}}$} &
  {$3\Om$} &
  {$\boxed{\bf10^{-18}}$} &
  {$\boxed{\bf10^{-20}}$} &
  $\om_L-3\Om$
   \\ 
$\gt_{ZY} $&
  $10^{-18}$ &
  {-} &
  -&
  {$\boxed{\bf10^{-18}}$} &
  {$\boxed{\bf10^{-19}}$} &
  {$3\Om$} &
  {$\boxed{\bf10^{-19}}$} &
  {$\boxed{\bf10^{-20}}$} &
  {$3\Om$}
   \\ 
$\gt_{DX} $&
  {$10^{-22}$} &
  - &
  - &
  $10^{-27}$ &
  - &
  - &
  $10^{-28}$ &
  - &
  -
   \\ 
$\gt_{DY} $&
  {$10^{-22}$} &
  - &
  - &
  $10^{-27}$ &
  - &
  - &
  $10^{-28}$ &
  - &
  -
   \\ 
$\gt_{DZ}$ &
  $10^{-22}$ &
  {-} &
  -&
  {$\boxed{\bf10^{-23}}$} &
  - &
  - &
  {$\boxed{\bf10^{-24}}$} &
  -  &
  -
  \\ \hline
\end{tabular}

\caption{The matter sector summary table reproduced from Ref.~\cite{datatable}
  with the omission of $c_\mn$ and the addition of the results of this Letter.
  The ``Current" columns list new limits obtained in this work from published results of searches at the sidereal frequency alongside existing limits in Ref.~\cite{datatable}. The ``Feasible" columns list estimated sensitivities that could be achieved in a suitable analysis conducted at the suggested frequency.
  Boxed results are new limits on coefficients that have not yet been constrained.
  Boldface indicates an improvement from the current best bounds in Ref.~\cite{datatable}. All numeric values reported are in units of GeV.
}
\label{datatables}

\end{table*}

Turning to a constraint-based perspective on this line of thinking,
if Lorentz violation is not seen,
we can ask how many Lorentz-violating degrees of freedom in a coefficient separation approach can be simultaneously probed using $H_b$ experiments and the powers of the boost considered here.
Of the 132 degrees of freedom observable with nonrelativistic ordinary matter, we excluded the 27 that are associated with $c_\mn$, which are not relevant in the context of $H_b$ experiments.
This leaves 35 degrees of freedom per fermion. 
By considering the dimension of the null space of distinct sets of measurements that are possible within our analysis,
we find that independent sensitivities to all 35 tilde coefficients can be achieved when a suitable set of 
$H_b$ experiments are performed at different latitudes and/or extract the amplitudes of the signals from a variety of the Fourier components associated with the frequencies in Eq.~\rf{freq}. It is also worth noting that which set of frequencies and latitudes is sufficient to constrain the entire space is dependent on the maximum order of magnitude chosen for the series expansion.

By including up to 24 orders of suppression, using two experiments at different latitudes involving the same species extracting signals at both the sine and cosine phases of the frequencies $\om$, $2\om$, $\Om$, $2\Om$, $3\Om$ for both horizontal laboratory directions suffices to span the full space of 35 coefficients. 
This same outcome can also be achieved using a single experiment, extracting signals at the same phases and directions of 12 different frequencies instead of five. One example of a suitable set of frequencies is 
$\om$, $2\om$, $\Om$, $2\Om$, $3\Om$, $\om+2\Om$, $\om-2\Om$, $\om+3\Om$, $\om-3\Om$, $2\om+2\Om$, $2\om-2\Om$, $2\om-3\Om$, 
but not all combinations of 12 frequencies are necessarily sufficient.
For experiments capable of extracting both phases of both horizontal laboratory directions, see Refs.~\cite{expdata:neutron-princeton,expdata:electron-washington}.

{\it Conclusion---}
We perform an analysis that obtains sensitivities to all of the remaining unmeasured coefficients in the minimal matter sector of the SME based on a reinterpretation of published results.  The complete coverage of this coefficient space is a significant milestone given the dozens of experiments that have been done since this sector of the SME was first developed more than 25 years ago. These maximum-reach results restrict the space available to model builders and allow experimenters to refocus their energies on other opportunities to search for Lorentz violation. We also provide estimates for numerous significant improvements that could be obtained from additional analysis of existing experimental datasets or the analysis of feasible new experiments.  Such analyses would offer added discovery potential.  This potential is further enhanced though our coefficient-separation result
in which we find that all of the spin-coupled Lorentz-violating degrees of freedom in the minimal SME can be simultaneously constrained using the methods introduced in this Letter.  Thus, the consideration of boost effects beyond first order can have a variety of significant impacts on the ongoing search for Lorentz violation as a signal of Planck-scale physics in the minimal fermion sector.  Moreover, the application of this approach in other sectors, as well as in the context of operators beyond the minimal level, offers a considerable space for further exploration.

{\it Acknowledgments--- }
This work was supported in part by the Ford Research Fund, Mathews Student Fellowship, and the Towsley Endowment at Carleton College.

{\it Data availability--- }
There are no publicly available research data or software supporting this manuscript. Requests
for further information or data should be sent to the authors.

\bibliography{refs}

\end{document}